\documentclass{PoS}
\usepackage{times,mathptm}
\usepackage{subfigure}
\usepackage{amsfonts,amsthm,bm,bbm,amsmath,amssymb,marvosym} 
\newcommand{\beq}{\begin{equation}}
\newcommand{\eeq}{\end{equation}}
\newcommand{\bea}{\begin{eqnarray}}
\newcommand{\eea}{\end{eqnarray}}
\renewcommand{\d}{\delta}
\renewcommand{\l}{\lambda}

\renewcommand{\b}{\beta}
\renewcommand{\a}{\alpha}

\newcommand{\vphi}{\varphi}
\newcommand{\tD}{\widetilde{D}}

\newcommand{\m}{\mu}
\newcommand{\bx}{{\bf x}}
\newcommand{\by}{{\bf y}}
\renewcommand{\r}{\rho}

\newcommand{\s}{\sigma}

\newcommand{\D}{\Delta}

\renewcommand{\th}{\theta}

\newcommand{\vph}{\varphi}
\newcommand{\oh}{\frac{1}{2}}

\newcommand{\dg}{\dagger}
\newcommand{\non}{\nonumber}

\newcommand{\rf}[1]{(\ref{#1})}
\newcommand{\ra}{\rightarrow}

\title{Aspects of Confinement in Coulomb Gauge}

\ShortTitle{Aspects of Confinement in Coulomb Gauge}

\author{\speaker{Jeff Greensite}%
        \thanks{Research supported in part by the U.S.\ Department of Energy under Grant No.\ DE-FG03-92ER40711.}\\
       Physics and Astronomy Department, San Francisco State University\\
       San Francisco, CA 94132 USA\\ \\
       E-mail: \email{jgreensite@gmail.com}}

\abstract{I present some new results regarding confinement as it appears in Coulomb gauge.  It is found that:  i) a recently proposed Yang-Mills vacuum wavefunctional in temporal gauge and 2+1 dimensions yields a Coulomb-gauge ghost propagator and linear Coulomb potential in good agreement with lattice Monte Carlo results;  ii) adding a few constituent gluons to heavy quark-antiquark states brings the interaction energy much closer to that of the static quark potential, and suggests the beginnings of  gluon-chain formation at roughly one fermi;  iii)  a perturbative approach to Faddeev-Popov eigenvalues indicates that the zero eigenvalue at the Gribov horizon may occur either at, or away from, $p=0$, depending on the gauge choice and spacetime dimension.   This last result may be relevant to the qualitatively different infrared behavior of the ghost propagator in Coulomb and Landau gauges.}

\FullConference{International Workshop on QCD Green's Functions, Confinement and Phenomenology\\
                September 7-11 , 2009\\
                ECT Trento, Italy}

\begin{document}

\section{Introduction}

    In this contribution I would like to report on three recent results concerning the non-perturbative behavior, in particular the confinement property, of Yang-Mills theory in Coulomb gauge.  The first result, regarding the Yang-Mills vacuum wavefunctional and the Coulomb potential that can be derived from it, and the second, concerned with the energetics of physical states
containing constituent gluons and a static quark-antiquark pair, were obtained in collaboration with {\v S}tefan Olejn\'{\i}k.   The latter work is presented in more detail in ref.\ \cite{Us1}.  In the last section I will outline another new result, concerning the spectra of the Faddeev-Popov operator at the Gribov horizon in less than four dimensions, in Coulomb and Landau gauges.
    
\section{The Vacuum Wavefunctional and the Coulomb Potential}

    A long time ago it was argued \cite{Me1} that the pure Yang-Mills vacuum wavefunctional, in temporal gauge, has  the following form at large distance scales:\footnote{A similar proposal was made by Halpern in 2+1 dimensions \cite{Marty}.}
\beq
            \Psi_0^{eff}[A] \approx \exp\left[-\m \int d^3 x ~ F_{ij}^a F_{ij}^a \right]
\label{dimred0}
\eeq
This vacuum state has the property of dimensional reduction, in the sense that computation of a large spacelike loop in 3+1 dimensions reduces to the calculation of a Wilson loop in Yang-Mills theory in 3 Euclidean dimensions.   Assuming the vacuum state $\Psi^{(2)}_{0}$ of the 2+1 dimensional theory also has the dimensional reduction property, the calculation of a planar spacelike loop in 3+1 dimensions reduces to a calculation in D=2 dimensions, i.e.
\bea
           W(C) &=& \langle \mbox{Tr}[U(C)]\rangle^{D=4} 
                      = \langle \Psi^{(3)}_{0}|\mbox{Tr}[U(C)]|\Psi_0^{(3)}\rangle
\non \\
                        &\sim&  \langle \mbox{Tr}[U(C)]\rangle^{D=3} 
                      = \langle \Psi^{(2)}_{0}|\mbox{Tr}[U(C)]|\Psi_0^{(2)}\rangle
\non \\
                        &\sim& \langle \mbox{Tr}[U(C)]\rangle^{D=2} 
\label{dimred1}
\eea
In D=2 dimensions the Wilson loop can be calculated analytically, and we know there is an area-law falloff.  Dimensional reduction therefore implies
confinement.

    Support for the dimensional reduction form of the vacuum wavefunctional comes from strong coupling expansions, and from numerical simulations.   
At strong couplings, in temporal gauge, the Yang-Mills ground-state has the form \cite{Me2}
\beq
\Psi_0[U] = \exp\left[{c\over g^4} \sum_{plaq} \mbox{Tr}[UUU^\dagger U^\dagger] + O(g^6) \right]
\label{dimred2}
\eeq
which clearly has the dimensional reduction property.  At weaker couplings, a numerical approach is available \cite{Junichi}.  We consider a modified lattice Monte Carlo simulation in temporal gauge, in which the configurations on the $t=0$ time-slice are restricted to belong to a finite set $\{U_\m^{(m)}({\bf x}),m=1,2,...,M\}$.  After each sweep through the rest of the lattice, the configuration at $t=0$ is chosen from among the given set via the Metropolis algorithm.  Let $N_m$ be the number of times the $m$-th configuration is selected.  Then it is easy to show that in the limit that the number of sweeps
$N_T \ra \infty$
\beq
         \left| {\Psi_0[U^{(m)}] \over \Psi_0[U^{(n)}]}\right|^2 = \lim_{N_T \ra \infty} {N_m\over N_n}   
\eeq
For ``large-scale'' configurations, e.g. non-abelian constant lattices and long-wavelength plane waves, the numerical results agree perfectly with the dimensional reduction form \rf{dimred0} \cite{Junichi}.

    Of course, the true vacuum can't be simply the dimensional reduction form; that would give incorrect results at short distances and high frequencies.
{\v S}tefan Olejn\'{\i}k and I have proposed, in ref.\ \cite{Us2}, that the ground state solution in D=2+1 dimensions, in temporal gauge, is approximated by 
\beq
\Psi_0[A] = \exp\left[-\oh\int d^2x d^2y ~ B^a(x) \left({1 \over \sqrt{-D^2 - \l_0 + m^2}} \right)^{ab}_{xy} B^b(y) \right]
\label{proposal}
\eeq
where $B^a=F_{12}^a$, $D^2$ is the covariant Laplacian in adjoint representation, $\l_0$ is the lowest eigenvalue of $-D^2$, and $m$ is a free parameter chosen from a fit to the string tension.  The motivations and successes of this wavefunctional, and the method which was developed to simulate it numerically, are described in the cited
reference (and in {\v S}tefan Olejn\'{\i}k's contribution to this meeting \cite{Stefan}).    Now the ground state wavefunctional in Coulomb gauge is simply the restriction of the temporal gauge wavefunctional to configurations satisfying the Coulomb gauge condition, i.e.\ $\Psi_0^{coul}[A] = \Psi_0[A]$  with $\nabla \cdot A = 0$.   If we denote by $g$ the gauge transformation taking a configuration $A$ to Coulomb gauge, then it is not hard to see that for any observable $Q[A]$, 
\beq 
\langle \Psi^{coul}_0|Q[A]|\Psi^{coul}_0 \rangle = \langle \Psi_0|Q[g\circ A]|\Psi_0 \rangle
\eeq
Then $\langle Q \rangle$ can be evaluated by generating lattices with probability weighting $\Psi_0^2$, transforming these lattices to Coulomb gauge, and evaluating the observable. 
Lattice configurations which are generated stochastically from the (latticized version of) $\Psi_0$ in eq.\ \rf{proposal} will be referred to as ``recursion lattices''.  The result for $\langle Q \rangle$ derived from recursion lattices can be compared with that obtained by generating lattices (which we refer to as ``MC lattices'') by the usual Monte Carlo procedure with a Wilson action, transforming to Coulomb gauge, and again evaluating the observable.  This latter procedure is the standard method for obtaining the expectation values of observables in Coulomb gauge.

\begin{figure}[htb]
\centerline{\scalebox{0.70}{\includegraphics{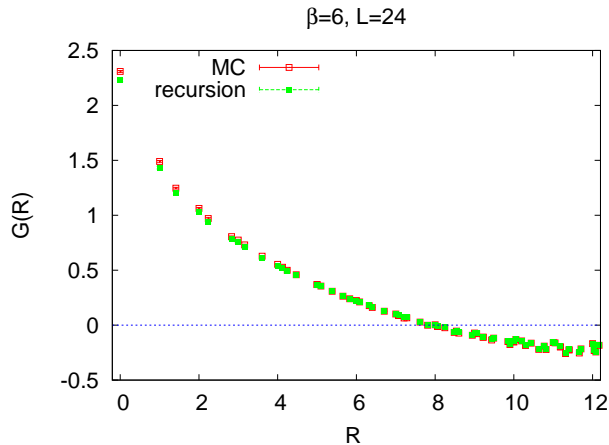}}}
\caption{The Coulomb ghost propagator evaluated on both recursion and MC lattices, at
lattice coupling $\b=6$ in $D=3$ dimensions.}
\label{gprop} 
\end{figure}

\begin{figure}[htb]
\centerline{\scalebox{0.70}{\includegraphics{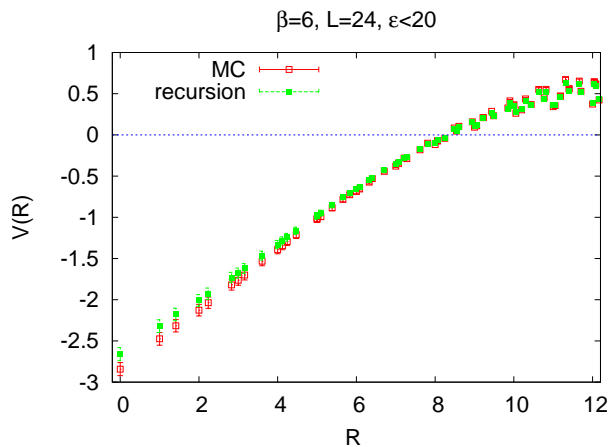}}}
\caption{Coulomb potential evaluated from data sets with a cut on lattices with $V(0)<-20$, for $\b=6$ and lattice extension $L=24$.  Results (with the same cut) are shown for both MC and recursion lattices.}
\label{Vcdat}
\end{figure}

    We have computed both the ghost propagator $G(R)$ and the Coulomb potential $V_C(R)$, defined by
\beq
G(R) = \left\langle - \left({1 \over \nabla \cdot D}\right)_{\bx \by}^{aa} \right\rangle ~~~,~~~ 
V_C(R) =  \left\langle \left({1 \over \nabla \cdot D} (-\nabla^2){1 \over \nabla \cdot D} \right)_{\bx \by}^{aa} \right\rangle 
\eeq
The result for the ghost propagator, obtained from both recursion and MC lattices, is shown in Fig.\ \ref{gprop}.  The Coulomb potential is very sensitive to ``exceptional'' configurations with very small eigenvalues of the Faddeev-Popov operator $-\nabla \cdot D$; these lead to huge errorbars.  To compare recursion and MC results, we impose cuts on the data, throwing away these rare configurations.  The result is shown in Fig.\ \ref{Vcdat}.
Obviously, the results for the ghost propagator and Coulomb potential obtained from our proposed wavefunctional closely agree with those obtained by standard methods.

\section{Constituent Gluons and the Gluon Chain Model}

The color Coulomb potential is known to be linear from computer simulations \cite{Us3}, but there are (at least) two serious difficulties in claiming that the Coulomb potential explains confinement.  First of all, the Coulomb string tension $\s_c$  is about three times larger than the asymptotic string tension $\s$.  Secondly, since the Coulomb force is essentially a one-gluon exchange effect, there are inevitably long-range dipole forces, which would result in long-range van der Waals forces among hadrons.  This latter problem is generic to any model of confinement based on ladder diagram exchanges,
and the difficulty can be traced to the absence of a flux tube.  This raises the question:  given that we already have a confining Coulomb potential, how and why does a flux tube form in Coulomb gauge?

\begin{figure}[htb]
\centering
\includegraphics[scale=0.45]{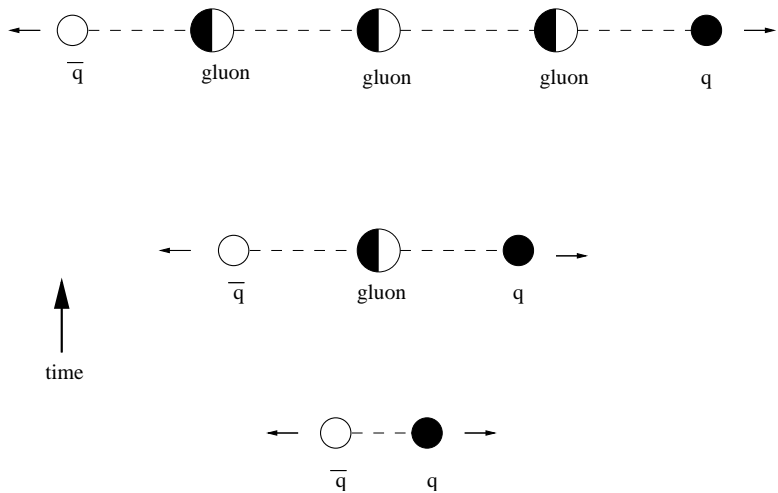}
\caption{The gluon chain model}
\label{gchain}
\end{figure}

    Let's begin with the fact that the Coulomb potential $V_C(R)$  is the interaction energy of a certain physical state containing two static charges, namely
\beq
\Psi_{q \overline{q}} = \overline{q}^a(0) q^a(R) \Psi_0 
\eeq
But this state is not necessarily the minimum energy state with two static charges, so the question is whether (and by how much) we can bring down the interaction energy, by adding some ``constituent'' gluons corresponding to acting on the vacuum with gluon field operators.  Schematically, these produce physical states of the form
\beq
\Psi_{q \overline{q}} = \overline{q}^a(0)\Bigl\{ c_0 + c_1 A + c_2 AA + ...\Bigr\}^{ab} q^b(R) \Psi_0 
\eeq
According to a scenario known as the ``gluon chain model'' (cf.\ ref.\ \cite{GT} and references therein, and ref.\ \cite{Tik})  the QCD color electric flux tube can be thought of, at least in certain gauges, as
a chain of constituent gluons.  The idea is that as a quark-antiquark pair separate, the electric field energy rises until a point is reached where it becomes energetically favorable to have a gluon in between the two quarks, thereby reducing the effective color charge separation.  As the quarks continue to separate, this process is repeated, and eventually the state resembles a chain of constituent gluons (Fig.\ \ref{gchain}), with the ordering of gluons in the matrix product of color indices correlated with the spatial ordering of gluons between the color charges.  This scenario has a number of attractive features.    A gluon chain has string-like properties (e.g.\ a Luscher term), Casimir scaling is obvious at large N, and the model is also consistent with N-ality dependence, in that string-breaking is natural for quark-antiquark sources in higher-dimensional color representations, at sufficient separations \cite{GT}.

    We would like to know if constituent gluons do indeed lower the interaction energy, for quark-antiquark separations that can be achieved in numerical lattice simulations.  Let us define, on the lattice, the rescaled transfer matrix
\beq
T = \exp[-(H-E_0)a]
\eeq
where $E_0$ is the energy of the ground state, $a$ is the lattice spacing, and $\exp(-Ha)$ is the usual transfer matrix.  Ideally, we would like to diagonalize $T$ in the subspace of states containing two static charges.  In practice, we must diagonalize in a finite $M$-dimensional subspace.   Let
\beq
 |k \rangle =  \overline{q}^a(\bx) Q_k^{ab}  q^b(\by) |\Psi_0 \rangle  ~~~~k=1,2,..,M
\eeq
where the $Q_k$ are functionals of the link variables.  In general the $\{|k\rangle\}$ are not orthogonal.  We then use standard lattice Monte Carlo methods to compute the following quantities
\bea
       O_{mn} &=& \langle m | n \rangle 
\non \\ &=& \langle \oh \mbox{Tr} [Q^\dg_m(t) Q_n(t) ] \rangle
\non \\ \\
       t_{mn}  &=&  \langle m |T| n \rangle 
\non \\
  &=&  \langle \oh \mbox{Tr} [Q^\dg_m(t+1) U^\dg_0(\bx,t) Q_n(t) U_0(\by,t) ] \rangle
\eea
From these quantities we can construct, via the Gram-Schmidt procedure, an orthonormal set of states $\{|\vphi_k\rangle\}$, and
and also derive the matrix elements $T_{ij} = \langle \varphi_i | T | \varphi_j \rangle$.  The next step is to numerically diagonalize the $T$-matrix in this finite basis, and then take
\beq
        V(R) =   - \log(\l_{max})
\eeq
where $\l_{max}$ is the largest eigenvalue of the $T$-matrix, as our estimate of the static quark potential.  

     We build the $Q_k$ operators out of ``smoothed'' A-fields.  Define, in the usual way, $A_\m(x) = (U_\m(x)-U^\dg_\m(x))/2i$.  These fields are
Fourier transformed, and the high-momentum components in directions transverse to direction $\mathbf{e}_j$ (direction of the $\bx_j$ axis) are 
exponentially suppressed according to
\bea
 A_i({\bf k},t) &\ra& \exp\Bigl[-\rho ({\bf k}^2 - k_j^2)\Bigr] A_i({\bf k},t)
\non \\
&\ra& \exp\Bigl[-\rho {\bf k}_\perp^2 \Bigr] A_i({\bf k},t)
\eea
where $\rho$ is a variational parameter.   Transforming back  to position space, we denote the resulting ``transverse-smoothed'' operator
$A_i(\bx,t,j)$ as the A-field smoothed in directions transverse to direction $\mathbf{e}_j$.  We also define $B_i(\bx,t) = 1 - \oh \mbox{Tr}[U_i(\bx,t)]$
and smooth in the same way to obtain  the transverse-smoothed operator $B_i(\bx,t,j)$ .
The $Q$ operators are then defined in terms of the $A_i(\bx,t,j)$ and $B_i(\bx,t,j)$,
for an antiquark at site  $\bx_0$  and a quark at site  $\bx_0 + R\mathbf{e}_j$                
\bea
Q_1(t) &=& \mathbbm{1}_2
\non \\
Q_2(t) &=& \sum_{n=0}^{R-1} A_j(\bx_0+n \mathbf{e}_j,t,j)
\non \\  
Q_3(t) &=& \sum_{n=-2}^{R+1} ~ \sum_{n'=n}^{R+1} A_j(\bx_0+n \mathbf{e}_j,t,j) A_j(\bx_0+n' \mathbf{e}_j,t,j)
\non \\
Q_4(t) &=& \sum_{n=-2}^{R+2} ~ \sum_{n'=n}^{R+2} ~ \sum_{i\ne j}
         \overline{A}_i(\bx_0+n \mathbf{e}_j,t,j) \overline{A}_i(\bx_0+n' \mathbf{e}_j,t,j)
\non \\
 Q_5(t) &=& \sum_{n=0}^{R-1} B_j(\bx_0+n \mathbf{e}_j,t,1) \mathbbm{1}_2
\non \\
 Q_6(t) &=& \sum_{n=0}^{R-1} \sum_{i\ne j} \overline{B}_i(\bx_0+n \mathbf{e}_j,t,j) \mathbbm{1}_2
\eea 
where we have also defined, for $i\ne j$, $\overline{A}_i(\bx,t,j) = \oh (A_i(\bx,t,j)+A_i(\bx-\mathbf{e}_i,t,j))$.
From these operators we construct and diagonalize the rescaled transfer matrix, as described above, in a truncated basis of six states.
We choose the variational parameter $\r$ which maximizes, at each quark-antiquark separation, the largest eigenvalue $\l_{max}$ of $T$.
Denote the corresponding eigenmode
\beq
          |\psi(R) \rangle = \sum_{n=1}^6 a_n(R) |\vphi_n\rangle
\eeq
Then the fraction of the norm of $\psi$ due to zero, one, and two or more constituent gluons is $|a_1|^2,|a_2|^2$, and $1-|a_1|^2-|a_2|^2$
respectively.

\begin{figure}[htb] 
\centering
\includegraphics[scale=0.7]{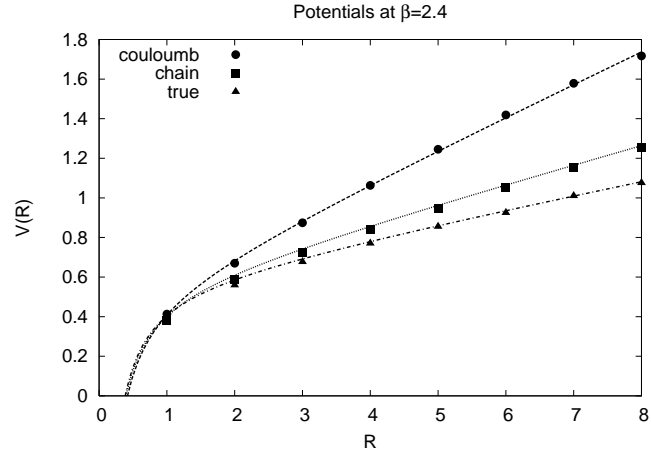}
\caption{The color Coulomb potential $V_{coul}(R)$, the ``gluon-chain'' potential $V(R)$ derived from the minimal-energy variational state, 
and the ``true'' static quark potential $V_{true}(R)$ obtained by standard methods. Results are shown at lattice coupling $\b=2.4$. 
Continuous lines are from a fit of data points.}  
\label{pots}
\end{figure}

\begin{figure}[htb]
\centering
\scalebox{0.7}{\includegraphics{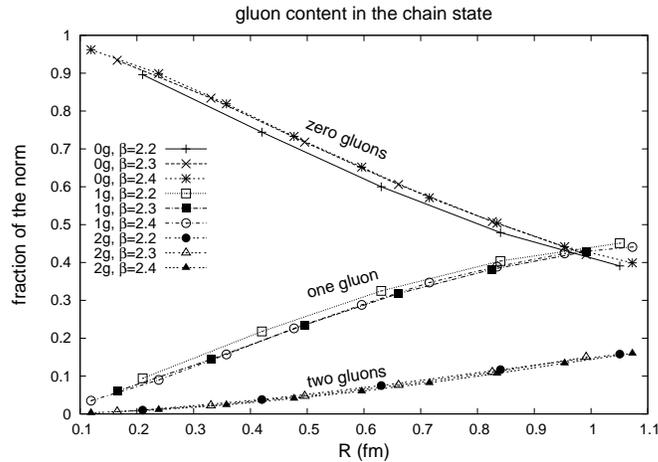}}
\caption{Zero, one, and two-gluon content (fraction of the norm of the variational state) vs.\
quark separation $R$ in fermis, at $\b=2.2,2.3,2.4$.}
\label{gluons}
\end{figure}

    Figure \ref{pots} shows our result for $V(R)$ at $\b=2.4$, together with the corresponding Coulomb potential (from $-\log(T_{11})$), and the static quark potential computed by the usual methods.  We see that i) constituent gluon operators bring the potential down from the Coulomb value to a value much closer to the usual static quark potential, and ii) the addition of constituent gluons does not affect linearity; the potential $V(R)$ of the minimal-energy variational state is still rising linearly.  Fig.\ \ref{gluons} shows the gluon content of the minimal energy state, where we see that at separations of about one fermi the variational state contains an equal admixture of zero and one-constituent gluon states.  Thus, at one fermi separation, we may be seeing the beginnings of a gluon-chain structure. This figure shows results obtained at $\b=2.2,2.3,2.4$, and is therefore also a test of scaling.

\begin{figure}[t!]
\centerline{\scalebox{0.7}{\includegraphics{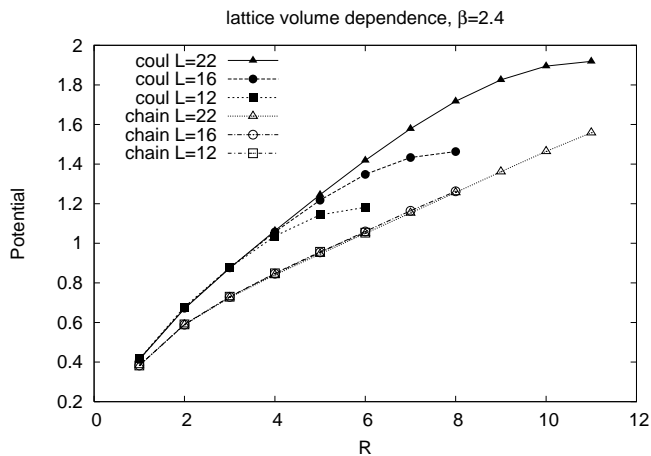}}}
\caption{Sensitivity of the Coulomb potential $V_{coul}(R)$ (solid symbols), and insensitivity of the chain potential $V_{chain}(R)$ (open symbols), to lattice volume.  Data is for the gauge coupling $\b=2.4$, and lattice volumes $L^4=12^4,16^4,22^4$.  Quark-antiquark separation $R$ is in lattice units.}
\label{coul24}
\end{figure}

    The dipole problem has already been mentioned.  The color Coulomb field is not expected to be collimated into a flux tube, and this means that there should be strong sensitivity to lattice volume, on a lattice of spatial extension $L$, for quark-antiquark separations close to $R=L/2$.
The reason is that for separations of that size, the finite volume cuts off a region where the field energy is still significant.  On the other hand, if
the field energy were collimated into a flux tube of diameter $d$, and if $L>>d$, then there would not be a similar sensitivity to the finite volume.
Figure \ref{coul24} displays the Coulombic and the variational (``chain'') potentials, computed on lattice volumes $12^4,16^4,22^4$. The results for variational, constituent-gluon states seem to be insensitive to lattice size, in contrast to the Coulomb potential.   This may be a hint that the dipole problem is much less severe for the constituent-gluon states.

\section{Faddeev-Popov spectrum at the Gribov horizon}

   ``Coulomb confinement'' means that the Coulomb energy of an isolated color charge is infinite.  An important question is whether this property follows from dominance, in the Coulomb gauge functional integral, of configurations lying on or near the first Gribov horizon \cite{Dan1}.   The first Gribov horizon is characterized by the existence of one non-trivial zero mode of the Faddeev-Popov (F-P) operator, and it is thought, since the ghost propagator is the inverse of the F-P operator, that an enhanced density of near-zero modes would result in a ghost dressing function which is singular at $p=0$.  It is a bit of a puzzle why this should be true in Coulomb gauge but not also in Landau gauge, where numerical simulations strongly suggest that the ghost dressing function is non-singular \cite{ghostprop}.   The question I would like to address is whether the near-zero modes associated with the first Gribov horizon necessarily correspond to singular behavior at $p=0$.
   
\begin{figure*}[tbh]
\begin{center}
\subfigure[] 
{
    \label{lowmass}
    \includegraphics[width=7truecm]{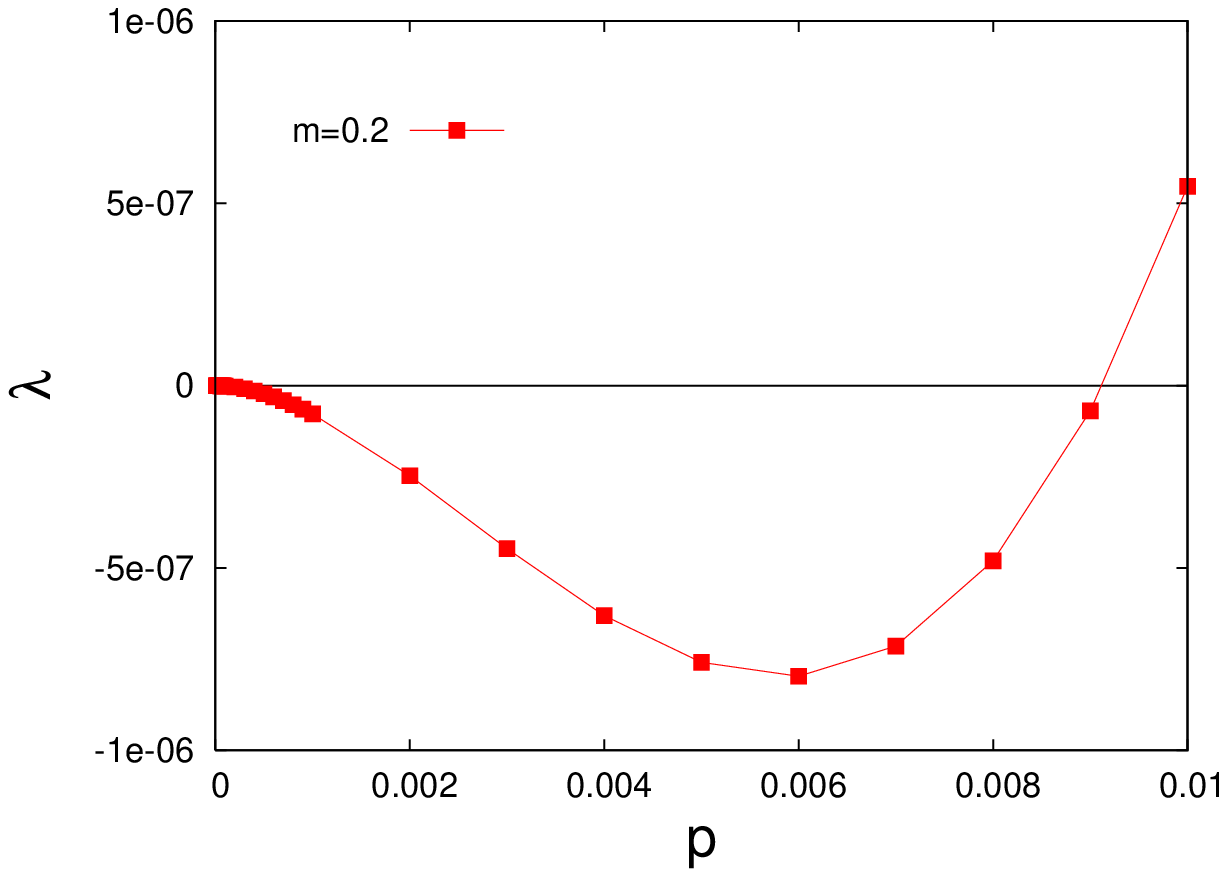}
}
\hspace{0.25cm}
\subfigure[] 
{
    \label{mcrit}
    \includegraphics[width=7truecm]{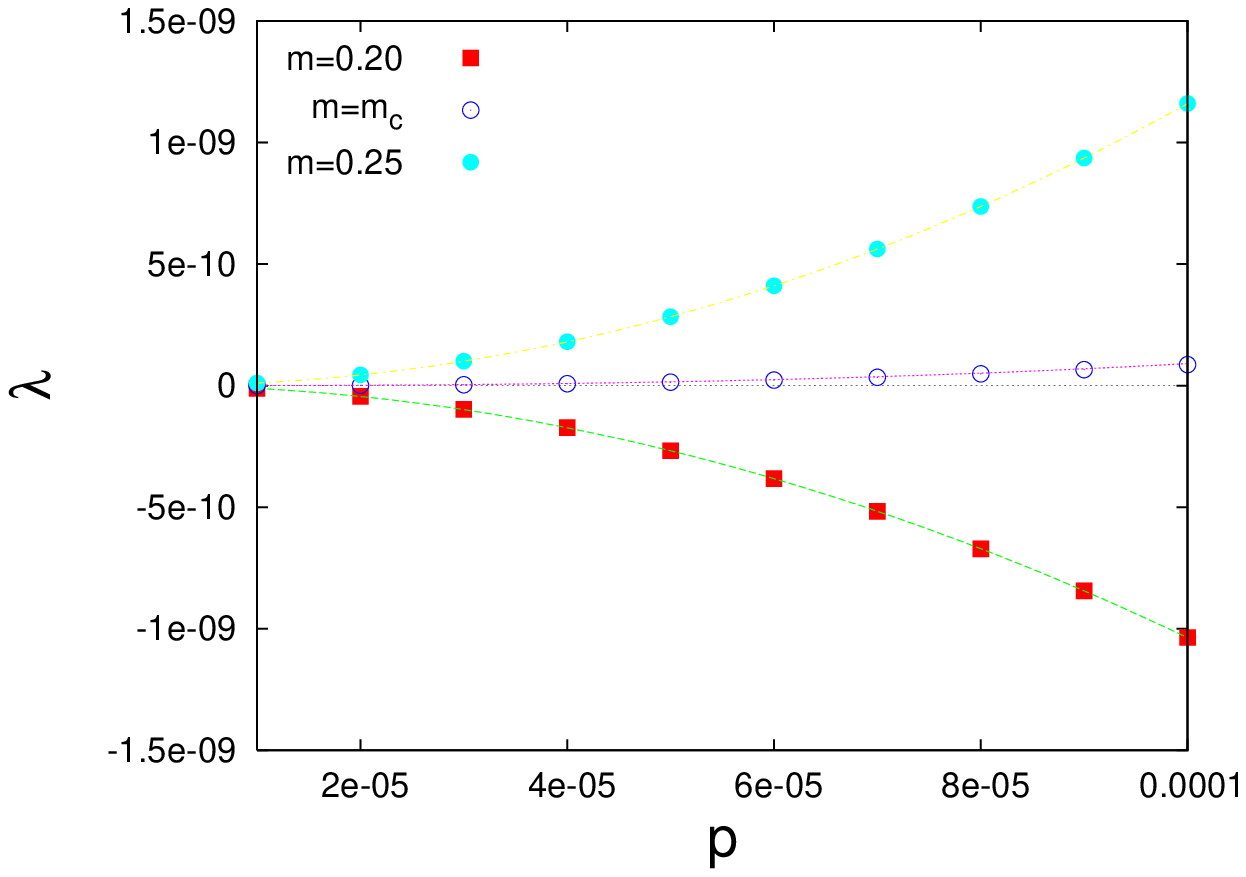}
}
\end{center}
\caption{F-P spectra at $\a=1$. (a) $m=0.20<m_c$.  There is an interval of negative eigenvalues in the region $0<p<0.009$. 
(b) $\l_p$ at low $p$, for $m$ below, above, and equal to the critical value $m_c=0.2228$.}  
\label{masses} 
\end{figure*}

      Ghost and gluon propagators have been studied extensively via Dyson-Schwinger equations;\footnote{This approach has a large literature.
There exist ``scaling'' solutions to the Dyson-Schwinger equation, with an infrared singular behavior for the ghost dressing function (cf.\ the review by Fischer \cite{Christian}), and there also exist ``decoupling" solutions \cite{decoupling} in which the ghost dressing function is finite.  At present, the numerical data in Landau gauge appears to favor the decoupling solution \cite{ghostprop}.} here I would like to investigate the spectrum of  the F-P operator via ordinary second-order perturbation theory, with non-perturbative information entering via an ansatz for the transverse gluon propagator. I work in less than four spacetime dimensions, to avoid the complications of renormalization. The F-P operator is
\bea
         M^{ac}  &=& -\delta^{ac} \nabla^2 - gf^{abc} A^b_i(x) \partial_i  
\nonumber \\ 
&=&  K_0 + g K_1
\eea 
with eigenvalue equation $M^{ab}\vphi^b_p = \l_p \vph_p^a$, and 
\beq
\langle \l_p \rangle = \l_p^{(0)} + \langle \D \l_p^{(1)} \rangle + \langle \D \l_p^{(2)} \rangle + ...
\eeq
with $\l_p^{(0)}=p^2$ at zeroth order, and  $\langle \D \l_p^{(1)} \rangle=0$ at first order in $g$.  The transverse gluon propagator is required for the second-order result, and I use $D^{ab}_{ij}(q) = \d^{ab}(\d_{ij} - q_i q_j/q^2)  D(q)$ with the ansatz for the dressing function
\bea
        D(q) = \left\{ \begin{array}{cl}
        {1 \over 2 \sqrt{q^2 + m^{2+\a}/q^\a}} & \mbox{Coulomb gauge} \cr \cr
       {1 \over q^2 + m^{2+\a}/q^\a} & \mbox{Landau gauge} \end{array} \right.
\eea
The second-order result in $d+1$ spacetime dimensions (Coulomb gauge), or $d$ spacetime dimensions (Landau gauge), is
\beq
    \langle \l_{p} \rangle = p^2\Bigl(1 - g^2 R_d I[p,m,\a]\Bigr)
\label{lp}
\eeq
where $R_d$ is a dimension-dependent constant of $O(1)$, and
\bea
I[p,m,\a]  &=&  \int_0^{\pi/2} d\th ~ \sin^{d-2}\th (1-\cos^2\th) 
\non \\
& & \qquad \times \left\{ \int_0^\infty dq 
{1 \over q +2p\cos\th}[\tD(4p\cos\th + q) + \tD(q)] \right.
\non \\
   & & \qquad \left. + \int_0^{2p\cos\th} dq {1 \over q - 2p\cos\th} [\tD(4p\cos\th - q) - \tD(q)] \right\}
\eea
with $\tD(q) = q^{d-2} D(q)$.  Eq.\ \rf{lp} is somewhat reminiscent of the Dyson-Schwinger equation for the ghost propagator.  However, this is an equation for the F-P eigenvalue spectrum, not the propagator, and of course there is no claim that the equation is exact.   In general, as $p\ra 0$,
\bea
 I[p,m,\a] &=& a[m,\a] - b[m,\a] p^s + ...
 \non \\
 \langle \l_p \rangle &=& (1-a[m,\a]) p^2 + b[m,\a] p^{2+s} + ... 
\eea
Then in Coulomb gauge, for fixed $\a$, I find three cases:
\beq
\begin{array}{cccc}
  1) & m<m_c, & a[m,\a] > 1, & \langle \l_p \rangle \sim - p^2 \cr \cr
  2) & m>m_c, & a[m,\a] < 1, &  \langle \l_p \rangle \sim + p^2 \cr \cr
  3) & m=m_c, & a[m,\a] = 1, &  ~~\langle \l_p \rangle \sim + p^{2+s} \cr
\end{array}
\eeq
When $m$ is below some critical mass parameter $m_c$, there is an interval of negative eigenvalues which begins at $p=0$. An example at $\a=1$, Coulomb gauge in 2+1 dimensions, is shown in Fig.\ \ref{lowmass}, with dimensional quantities in units of $g^2$.  This case corresponds to the dominant gluon field configurations lying outside the Gribov region.  As $m$ increases, the interval of negative eigenvalues shrinks in size, until at some critical $m=m_c$ the interval shrinks to one point.  This would correspond to the dominant field configurations lying right on the first Gribov horizon.  At $m>m_c$ we have $\l_p \sim p^2$, and the relevant gluon field configurations lie inside the Gribov region. The spectrum above, at, and below $m_c$, at near-zero values $p$, is shown in Fig.\ \ref{mcrit}, again for  $\a=1$ and Coulomb gauge in 2+1 dimensions. 

\begin{figure}[h!]
\centerline{\includegraphics[width=8truecm]{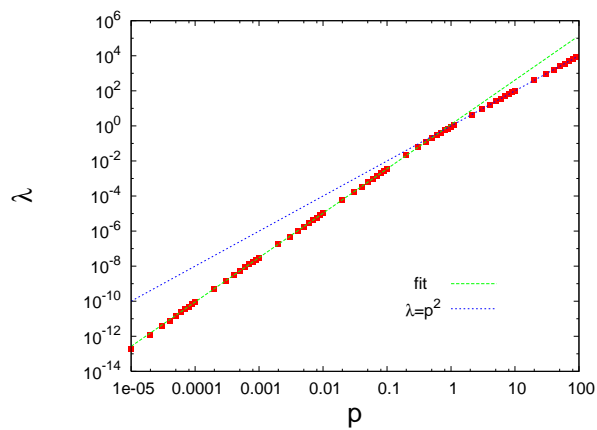}}
\caption{Log-log plot of the spectrum of the Faddeev-Popov operator (2+1 dimensional Coulomb gauge), for
$\a=1$ at the critical $m_c=0.223$. A best fit at $p<1$ yields $\l_p = 1.21 p^{2.53}$.}
\label{alf10}
\end{figure}

    The $m=m_c$ case is the one of interest, because this is where the near-zero F-P eigenvalues rise with a non-standard power $p^{2+s}$, leading
to an enhanced eigenvalue density and (it can be shown) Coulomb confinement in 2+1 dimensions for any $s>0$.  For any fixed $\a$ we can calculate $s$ and the critical gluon mass parameter $m_c$.  For example, at $\a=1$ it is found that $m_c=0.223$, and the exponent $2+s$ is determined from a straight-line fit to a log-log plot of $\l_p$ vs.\ $p$, as seen in Fig.\ \ref{alf10}.  From this plot, we  find $\l_p \sim p^{2.53}$ at small $p$ and $m=m_c$, going back to $\l_p \sim p^2$ at larger $p$.

\begin{figure}[tbh]
\centerline{\scalebox{0.7}{\includegraphics{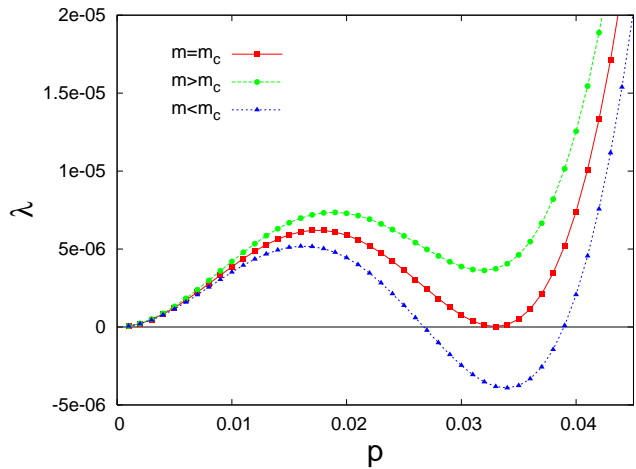}}}
\caption{The low-lying F-P spectrum in Landau gauge, in $D=3$ dimensions and $\a=1$, for gluon mass parameter $m$ above, below, and equal to the critical value $m_c$.}
\label{mcritL}
\end{figure}

    Landau gauge is a little different.  In two spacetime dimensions the F-P Landau gauge spectrum is qualitatively similar to the Coulomb gauge result in three spacetime dimensions.  However, in three spacetime dimensions, for any $\a>0$, it turns out that the interval of negative eigenvalues at $m<m_c$ is located \emph{away} from $p=0$, and at the critical $m=m_c$ the non-trivial zero eigenvalue is at some $p>0$.  This situation is shown in Fig.\ \ref{mcritL}.   

    While the spectrum of F-P eigenvalues does not translate directly into a prediction for the behavior of the ghost propagator (because the
momentum behavior of the F-P eigenmodes must also be taken into account), what we do see is that two different scenarios for the behavior of the eigenmode spectrum at the Gribov horizon are possible.  In the first scenario, the non-trivial zero eigenvalue is at $p=0$, while in the other, it is located away from $p=0$.  It is natural to conjecture that the first scenario is associated with the confining properties of the ghost propagator in Coulomb gauge, while the second scenario has something to do with non-singular infrared behavior of the ghost dressing function in Landau gauge.

\end{document}